\documentclass[prl,superscriptaddress,reprint,showpacs,longbibliography]{revtex4-2}
\usepackage{lineno}
\usepackage{color}
\usepackage{physics}
\usepackage{graphicx,textcomp,amssymb,amsmath,dcolumn,hyperref}
\usepackage{siunitx}
\usepackage{comment}

\begin{document}

\title{Pauli blockade catalogue and three- and four-particle Kondo effect in bilayer graphene quantum dots}

\author{Chuyao Tong}
\email{ctong@phys.ethz.ch}
\affiliation{Solid State Physics Laboratory, ETH Zurich, CH-8093 Zurich, Switzerland}

\author{Annika Kurzmann}
\affiliation{Solid State Physics Laboratory, ETH Zurich, CH-8093 Zurich, Switzerland}
\affiliation{2nd Institute of Physics, RWTH Aachen University, Aachen, 52074, Germany}

\author{Rebekka Garreis}
\affiliation{Solid State Physics Laboratory, ETH Zurich, CH-8093 Zurich, Switzerland}

\author{Kenji Watanabe}
\affiliation{Research Center for Functional Materials, National Institute for Materials Science, 1-1 Namiki, Tsukuba 305-0044, Japan}
\author{Takashi Taniguchi}
\affiliation{International Center for Materials Nanoarchitectonics, National Institute for Materials Science,  1-1 Namiki, Tsukuba 305-0044, Japan}

\author{Thomas Ihn}
\author{Klaus Ensslin}
\affiliation{Solid State Physics Laboratory, ETH Zurich, CH-8093 Zurich, Switzerland}

\date{\today}

\begin{abstract}
Pauli blockade is a fundamental quantum phenomenon that also serves as a powerful tool for qubit manipulation and read-out. While most systems exhibit a simple even-odd pattern of double-dot Pauli spin blockade due to the preferred singlet pairing of spins, the additional valley degree of freedom offered by bilayer graphene greatly alters this pattern. Inspecting bias-triangle measurements at double-dot charge degeneracies with up to four electrons in each dot reveals a much richer double-dot Pauli blockade catalogue with both spin and/or valley blockade. In addition, we use single-dot Kondo effect measurements to substantiate our understanding of the three- and four-particle state spectra by analyzing their magnetic field dependence. With high controllability and reported long valley- and spin-relaxation times, bilayer graphene is a rising platform for hosting semiconductor quantum dot qubits.
A thorough understanding of state spectra is crucial for qubit design and manipulation, and the rich Pauli blockade catalogue provides an abundance of novel qubit operational possibilities and opportunities to explore intriguing spin and valley physics.
\end{abstract}
\maketitle
$\mathcal{H}$
Carbon-based materials are promising hosts for spin-qubits~\cite{Trauzettel2007spinqubit,RMPspin, liu20192d,jing2022gate} due to the natural abundance of $98.9\%$ low-mass, nuclear-spin free $^{\mathrm{12}}$C. Experimental advancements achieved in recent years~\cite{MariusPRX, Eich2018coupled, banszerus2018doubledot, Kurzmann2019Detector, Tunabledot, RebekkaPRL,annikaexcitedstates, AnnikaKondo, Luca1001, Lucacrossover, Blockade, SpinT1, SpinT1Aachen,Samuel2particles,Leakage, BlockadeReadout,banszerus2023particle} have demonstrated the great potential for electrostatically defined bilayer graphene (BLG) quantum dots (QDs) to host spin and valley qubits, with highly tunable QDs' geometry and tunnel couplings, as well as gate-tunable valley $g$-factors~\cite{Tunabledot}, and reported remarkably long valley-relaxation time $T_1>\SI{500}{ms}$~\cite{BlockadeReadout}, and spin-relaxation times $T_1$ up to $\SI{50}{ms}$~\cite{SpinT1, SpinT1Aachen, BlockadeReadout}, comparable with the state-of-the-art results achieved in other semiconductor quantum dot systems~\cite{stano2021review}.

\begin{figure}
	\includegraphics[width=8.5cm]{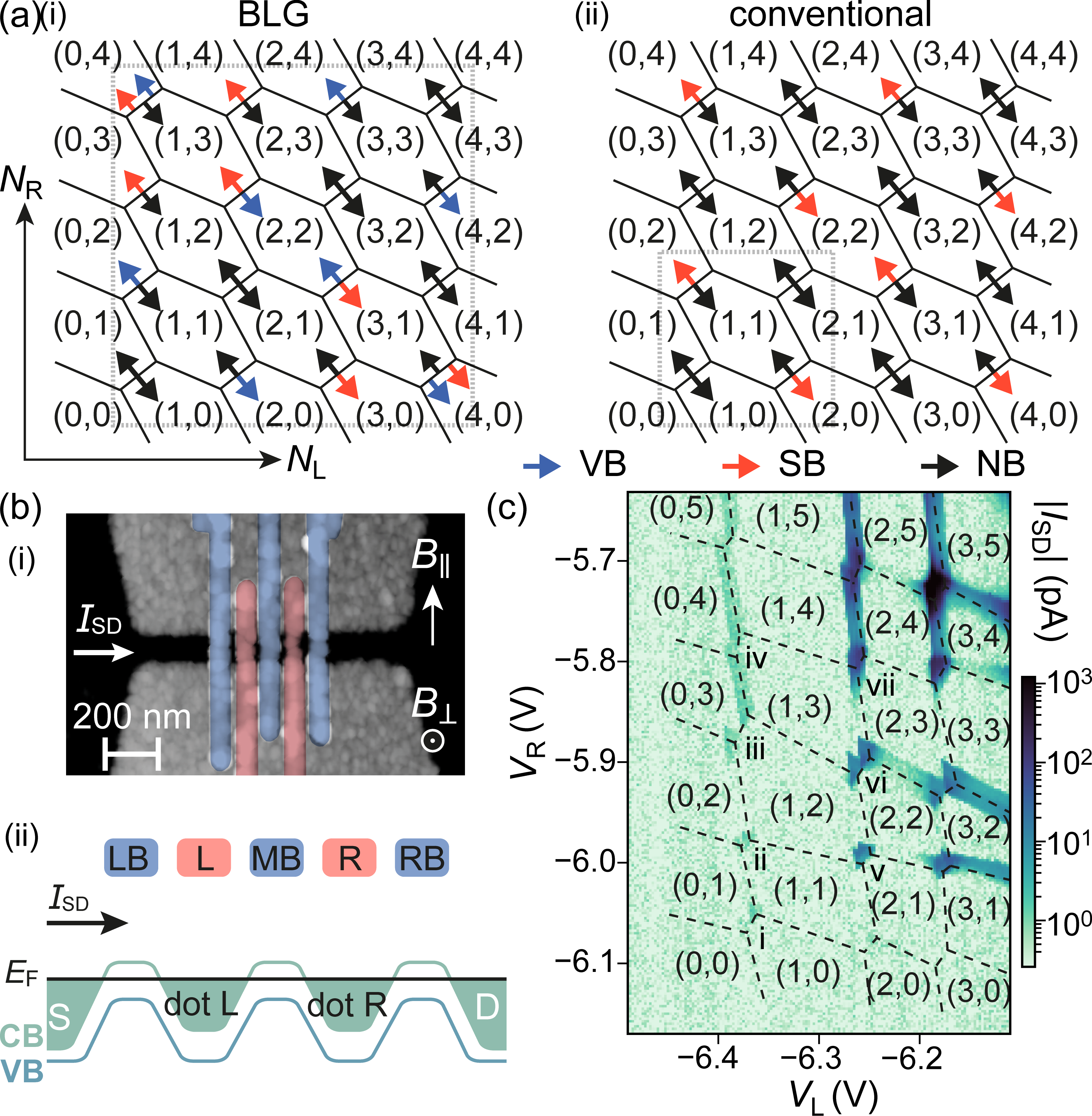}
	\caption{(a) Schematic double-dot charge stability diagram for (i) BLG, and (ii) a conventional system~\cite{JohnsonBlockade}. Non-blocked (NB), spin-blocked (SB), and valley-blocked (VB) transitions are marked with black, red, and blue arrows. (b)(i) False-colored AFM image of the device. (ii) Conduction band (CB) and valence band (VB) edge variation along the channel. Dots L and R are formed underneath the respective plunger gates (red), with gate voltages $V_\mathrm{L}$ and $V_\mathrm{R}$. Left-barrier (LB), middle-barrier (MB), and right-barrier (RB) gates (blue) control the respective tunnel barriers. (c) Double-dot charge stability map at $V_\mathrm{SD}=\SI{1.5}{mV}$ with stable charge states marked in the respective Coulomb-blockaded regions and separated with dashed lines. Roman numerals label the triple points studied in this work, with zoom-ins presented in Fig.~\ref{fig2}(b). The number of electrons in the left $N_\mathrm{L}$ and in the right $N_\mathrm{R}$ dot increase from left to right, and from top to bottom. }
	\label{fig1}
\end{figure}

To harvest the full potential of BLG QDs, a thorough understanding of the relevant QD state spectra is essential. In BLG, in addition to spins, up and down, there exist two valleys, $\mathrm{K^+}$ and $\mathrm{K^-}$
, which couple to an external perpendicular magnetic field due to their non-trivial Berry curvature~\cite{Knothe2018, Moulsdale2020, MariusPRX, Tunabledot, RebekkaPRL}. 
These valleys enrich the spectra of BLG, and lead to intriguing properties, such as the spin-triplet valley-singlet 
single-dot two-particle ground state~\cite{annikaexcitedstates,Samuel2particles, AngelikaQuartetStates,Knothe2022}, and the double-dot two-carrier Pauli spin- and valley-blockade~\cite{Blockade}. 
Most other systems, i.e. systems without valleys, exhibit no Pauli blockade when an \emph{odd} number of carriers reside in a double dot due to the preferred singlet pairing of spins~\cite{ JohnsonBlockade, petta2005coherent, koppens2005control, johnson2005triplet,  PhysRevLett.103.160503, PhysRevLett.108.046808}. 
In contrast, in BLG three-carrier spin blockade has been demonstrated recently~\cite{Leakage}.  

Not limited to three carriers, 
the entire double-dot Pauli blockade catalogue with up to four carriers (i.e., a full shell) in each dot is far richer for BLG than for a conventional system. Our main result is summarized in Fig.~\ref{fig1}(a), a comparison of schematic double-dot charge stability diagrams for (i) BLG and for (ii) a conventional system.
In a conventional system, an even-odd pattern of Pauli spin-blocked and non-blocked transitions arise~\cite{JohnsonBlockade}; in contrast, in BLG 
Pauli blockade could arise from both spin \emph{and} valley selection rules. The rich blockade structure provides a multitude of novel qubit operational positions, allowing us to exploit the unique spin and valley physics for varied 
qubit manipulation and control~\cite{cai2023coherent}.

In this work, we study transport measurements through BLG double quantum dot (DQD). By analyzing finite-bias triangles at various charge degeneracies, we discuss the states and transitions involved, and hence the nature of the observed Pauli blockade as summarized in Fig.~\ref{fig1}(a), going beyond the previously discussed two-carrier~\cite{Blockade} and three-carrier~\cite{Leakage} cases.
Substantiating our understanding of the three- and four-particle state spectra experimentally, we further present transport measurements through a single BLG QD, probing the magnetic field dependence of the state spectra with the Kondo effect (similar to Ref.~\cite{AnnikaKondo}).

The BLG QDs are defined electrostatically [device shown in Fig.~\ref{fig1}(b)]
. We use the bandgap arising from a perpendicular displacement field~\cite{Ohta2006BLGband,mccann2006BLGband,Oostinga2008BLG}, achieved by the global backgate and various local top-gates. The split-gates [Fig.~\ref{fig1}(b,i), gray] form a 1D channel. Another layer of gates (red and blue) locally tune the potential landscape within this channel. See Supplemental Materials~S1~\cite{supp} for more detail. 

We form an electron DQD with $n$-type leads and tunable tunnel barriers [potential landscape depicted in Fig.~\ref{fig1}(b,ii)].
The plunger gate voltages $V_\mathrm{L}$ and $V_\mathrm{R}$ define dots L and R underneath the respective gates (red). More negative $V_\mathrm{L}$ and $V_\mathrm{R}$ deplete the respective dot discretely to the last electron. When a voltage $V_\mathrm{SD}$ is applied between the source and drain leads, finite-bias triangles form at triple-points where three charge configurations coexist. The distinctive honeycomb pattern of the double-dot charge stability diagram is observed on Fig.~\ref{fig1}(c). Charging energies $\sim$6--$\SI{10}{meV}$ are estimated for both dots, much larger than the energy scales that will be considered in the following discussions.

\begin{figure*}
	\includegraphics[width=17.8cm]{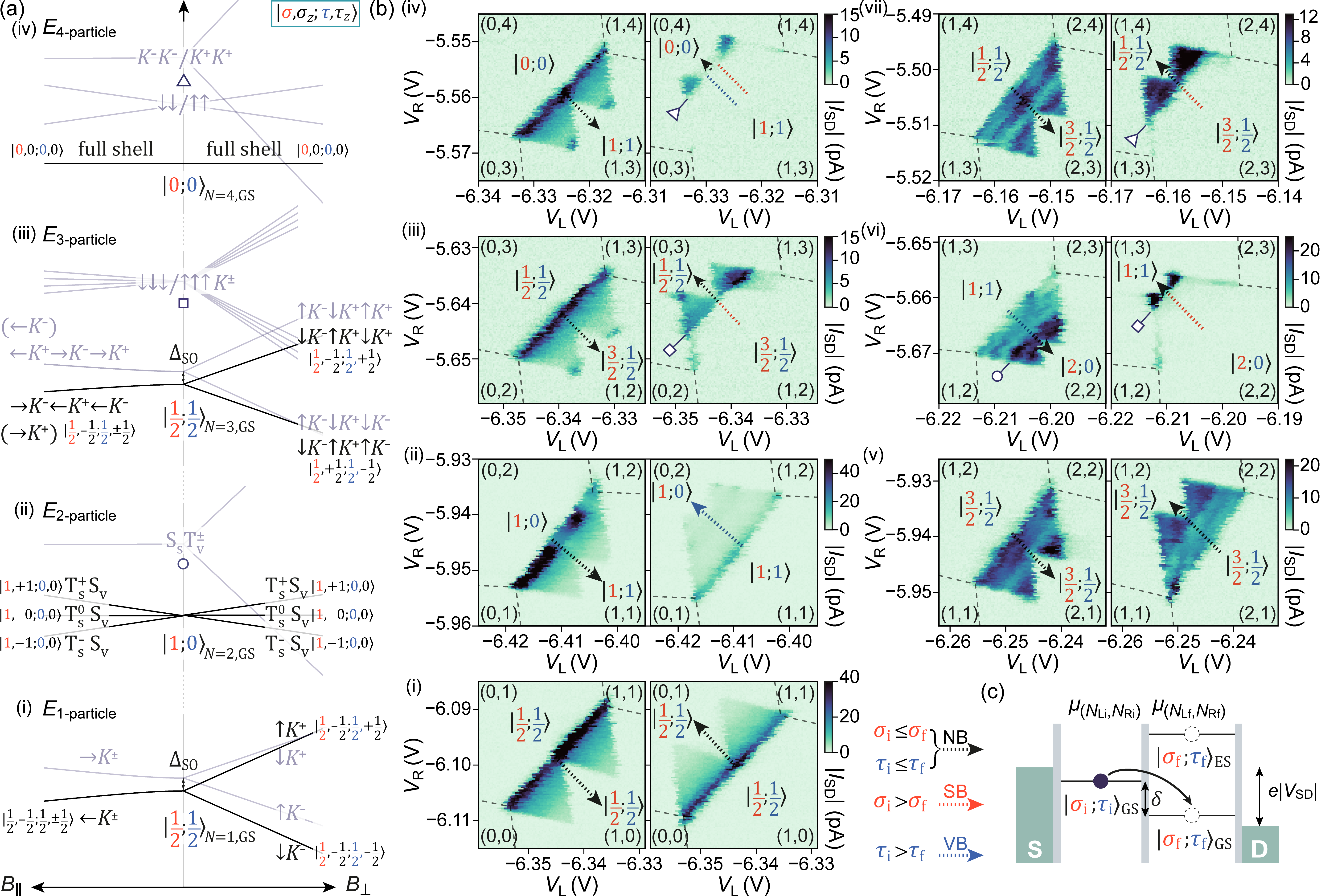}
	\caption{(a) $N$-particle state spectra $E_{N}$ with zero-field ground states (black) and relevant excited states (gray) in in-plane $B_\parallel$ and out-of-plane $B_\perp$ magnetic field, for (i)--(iv) $N=1$--$4$. The zero-field $N$-particle ground state is labeled as $\ket{\sigma;\tau}_{N,\mathrm{GS}}$ where $\sigma$ (red) and $\tau$ (blue) is its total spin and valley number. (b) Finite-bias triangles at $V_\mathrm{SD}=\SI{-1}{mV}$ (left panel) and $\SI{+1}{mV}$ (right panel), at charge degeneracies (i) $(0,1)$--$(1,0)$, (ii) $(0,2)$--$(1,1)$ (reproduced from supplementary information of Ref.~\cite{Blockade}), (iii) $(0,3)$--$(1,2)$ (reproduced from Ref.~\cite{Leakage}), (iv) $(0,4)$--$(1,3)$, (v) $(1,2)$--$(2,1)$, (vi) $(1,3)$--$(2,2)$, and (vii) $(1,4)$--$(2,3)$. Arrows mark the direction of charge configuration transitions; non-blocked (NB), spin-blocked (SB), and valley-blocked (VB) transitions are marked with black, red, and blue arrows, respectively. Ground states of each charge configuration are labeled as $\ket{\sigma=\sigma_\mathrm{NL}+\sigma_\mathrm{NR};\tau=\tau_\mathrm{NL}+\tau_\mathrm{NR}}$. (c) Schematic illustrating the condition of Pauli blockade, determined by the total spin (valley) number of the ground states of the initial $\sigma_\mathrm{i}$ ($\tau_\mathrm{i}$) and final $\sigma_\mathrm{f}$ ($\tau_\mathrm{f}$) charge configuration.}
	\label{fig2}
\end{figure*}

The nature of Pauli blockade is determined by the relevant spin and valley states involved. Understanding the blockade therefore demands a discussion on the single-dot $N$-particle state spectra [sketched in Fig.~\ref{fig2}(a)
]. We choose to discuss the states first with the single-particle notation:

\emph{One-particle}: 
The four-fold spin and valley states are split by a small (60--$\SI{80}{\micro eV}$) Kane--Mele~\cite{AnnikaKondo, Luca1001, KaneMele} spin--orbit gap $\Delta_\mathrm{SO}$ into two Kramer pairs: lower-energy $\downarrow\mathrm{K^-}$ and $\uparrow\mathrm{K^+}$, and higher-energy $\downarrow\mathrm{K^+}$ and $\mathrm{\uparrow K^-}$. At zero field, $\Delta_\mathrm{SO}$ quantizes the spins and valleys along the direction perpendicular to the BLG sheet.

\emph{Two-particles}: The spin-triplet valley-singlet is the ground state, as shown in Refs.~\cite{annikaexcitedstates,Samuel2particles,AngelikaQuartetStates,Knothe2022} and Fig.~\ref{fig2}(a,ii).

\emph{Three-particles}: Taking three of the four available states for one-particle gives us the spectrum shown in Fig.~\ref{fig2}(a,iii), with Kramers pairs $\mathrm{\downarrow K^-\uparrow K^+\downarrow K^+}$ and $\mathrm{\downarrow K^-\uparrow K^+\uparrow K^-}$ lower, and $\mathrm{\downarrow K^+\uparrow K^-\downarrow K^-}$ and $\mathrm{\downarrow K^+\uparrow K^-\uparrow K^+}$ higher in energy.

\emph{Four-particles}: Four carriers complete the shell, occupying all the available states in the lowest energy orbital.

Alternative to the single-particle notation, we can also write the states with their quantum numbers: the total spin (valley) number $\sigma$ ($\tau$), and their projections to the spin (valley) quantization-axis $\sigma_z$ ($\tau_z$). We define $\sigma_z=\pm1/2$ for spins $\uparrow/\downarrow$, and $\tau_z=\pm1/2$ for valleys $\mathrm{K^\pm}$. This notation $\ket{\sigma,\sigma_z;\tau,\tau_z}$ is marked in Fig.~\ref{fig2}(a) for the ground states at zero magnetic fields (black). The valley quantization-axis is always out-of-plane, whereas the spin quantization-axis is out-of-plane at zero $B$-field due to $\Delta_\mathrm{SO}$~\cite{AnnikaKondo, Luca1001}, and tilts towards in-plane in external $B_\parallel$. A $B$-field applied in any direction shifts spin states energies by $\sigma_{z}g_\mathrm{s}\mu_\mathrm{B}B$, while only an out-of-plane field shifts the valley states energies by $\tau_z g_\mathrm{v}\mu_\mathrm{B}B_\perp$, where $\mu_\mathrm{B}$ is the Bohr magneton, $g_\mathrm{s}=2$ the spin $g$-factor~\cite{MariusPRX,RebekkaPRL,annikaexcitedstates}, and $g_\mathrm{v}\gg g_\mathrm{s}$ the gap-size and dot-geometry dependent valley $g$-factor~\cite{AnnikaKondo,Tunabledot}. 

Since the interdot coupling is weak, we can choose to write the double-dot states as product states of the single dot states in the left and in the right dot, written in a bracket $(\ket{N_\mathrm{L}},\ket{N_\mathrm{R}})$. The spin (valley) projection of a double-dot state is thus simply the sum of its constituting single dot states $\sigma_z=\sigma_{z,\mathrm{NL}}+\sigma_{z,\mathrm{NR}}$ ($\tau_z=\tau_{z,\mathrm{NL}}+\tau_{z,\mathrm{NR}}$). Since both spin--orbit~\cite{KaneMele,BLGSOItheory} and valley--orbit~\cite{BLGVOItheory} effects are weak in BLG, we assume that tunneling conserves all the spin and valley numbers
. Transitions between double-dot charge configurations are \emph{not} blocked only if, for every degenerate ground state of the initial charge configuration, there exists a state in the final charge configuration being lower in energy and matching all quantum numbers with the initial state. Otherwise, some form of Pauli blockade arises; its nature depends on the mismatched quantum number (spin and/or valley).


In Fig.~\ref{fig2}(b), we present zoom-ins of the finite-bias triangles labeled with Roman numerals i--vii in Fig.~\ref{fig1}(c). The barrier gate voltages are adjusted slightly such that the dot geometry and tunnel couplings are similar for different charge transitions. 
Charge-configuration transitions (dashed arrows) can be non-blocked (black), spin-blocked (red), or valley-blocked (blue):

(i) \emph{$(1,0)$--$(0,1)$}: 
Any loaded electron hops between the two dots with the same one-particle states. No blockade is observed: the bias-triangles are complete for both bias directions.

(ii) \emph{$(1,1)$--$(0,2)$}: Studied in Ref.~\cite{Blockade},
the system is stuck when a $(1,1)$ state with $\tau_z=\pm1$ (valley-polarized state, e.g., $\mathrm{K^-}$ in both dots) is loaded, as transition to the $(0,2)$ ground state with $\tau_z=0$ (paired valley-singlet $\mathrm{S_v}$) is forbidden by valley conservation. The valley-blocked transition $(1,1)\rightarrow(0,2)$ suppresses the current in the bias-triangles [right, Fig.~\ref{fig2}(b,ii)], as compared to the non-blocked ones (left).  

(iii) \emph{$(1,2)$--$(0,3)$}: Studied in Ref.~\cite{Leakage},
transition $(1,2)\rightarrow(0,3)$ is spin-blocked: The system is stuck in the maximally spin-polarized $(1,2)$ states $\sigma_z=\pm3/2$ [e.g. $(\downarrow,\mathrm{T^-_s})$], since the $(0,3)$ ground state can only offer $\sigma_z=\pm1/2$ states~\cite{PhysRevB.82.075403, PhysRevB.95.035408}. Current in the $(1,2)\rightarrow(0,3)$ bias-triangles [right, Fig.~\ref{fig2}(b,iii)] is therefore strongly suppressed, as compared to the non-blocked $(0,3)\rightarrow(1,2)$ ones (left). 
The spin-blockade current suppression is stronger here than that of valley-blockade in (ii). 
Marked by a square at detuning $\sim\SI{0.6}{meV}$, transport resumes when the $(0,3)$ excited state with $\sigma_z=\pm3/2$ [$(0,\downarrow\downarrow\downarrow/\uparrow\uparrow\uparrow)$, Fig.~\ref{fig2}(a,iii) gray] becomes lower in energy than the $(1,2)$ ground state.

(iv) \emph{$(1,3)$--$(0,4)$}: Transition $(0,4)\rightarrow(1,3)$ is non-blocked, where $(1,3)\rightarrow(0,4)$ is both spin \emph{and} valley blocked with strong current suppression:
The $(1,3)$ ground state can be spin ($\sigma_z=\pm1$) or valley ($\tau_z=\pm1$) polarized, whereas $(0,4)$ only accepts a spin \emph{and} valley paired full shell. The blockade lifts (indicated by the triangle) when the polarized $(0,4)$ excited state with $\sigma_z=\pm1$ $(0,\downarrow\downarrow/\uparrow\uparrow)$ or $\tau_z\pm1$ $(0,\mathrm{K^-K^-/K^+K^+})$ [gray in Fig.~\ref{fig2}(a,iv)] becomes accessible at detuning $\sim\SI{0.8}{meV}$.

(v) \emph{$(2,1)$--$(1,2)$}: 
Since the states in the two charge configurations are essentially the same with only the roles of the left and the right dot switched, no blockade exists in either direction. 

(vi) \emph{$(2,2)$--$(1,3)$}:
Both sets of bias-triangles in Fig.~\ref{fig2}(b,vi) show regions with current suppression, with spin-blockade (right) stronger than valley-blockade (left). For $(2,2)\rightarrow(1,3)$, 
a fully spin-polarized $(2,2)$ ground state with $\sigma_z=\pm2$, e.g. $(\mathrm{T^-_sS_v},\mathrm{T^-_sS_v})$, is blocked to the $(1,3)$ ground state where the maximally spin-polarized state has only $\sigma_z=\pm1$, e.g. $\mathrm{(\downarrow K^-,\downarrow K^-\uparrow K^+\downarrow K^+)}$. This spin-blockade is lifted (marked by the square) upon accessing the spin-polarized excited three-particle states $\downarrow\downarrow\downarrow/\uparrow\uparrow\uparrow$ in the right dot [gray in Fig.~\ref{fig2}(a,iii)]. For $(1,3)\rightarrow(2,2)$, 
a valley-polarized $(1,3)$ ground state with $\tau_z=\pm1$, e.g. $(\mathrm{\downarrow K^-,\downarrow K^-\uparrow K^+\uparrow K^-})$ is blocked to the $(2,2)$ ground state that allows only fully-paired valleys $(\mathrm{T_sS_v},\mathrm{T_sS_v})$ with $\tau_z=0$. This valley blockade is lifted (marked by the circle) upon accessing the valley-polarized excited two-particle states $\mathrm{S_sT^\pm}$ [gray in Fig.~\ref{fig2}(a,ii)] in either dot. Here, transitions in both directions are Pauli blocked but for different quantum numbers---an intriguing situation special to BLG.

(vii) \emph{$(1,4)$--$(2,3)$}:
The transition $(2,3)\rightarrow(1,4)$ is spin-blocked and $(1,4)\rightarrow(2,3)$ non-blocked, demonstrated by the strong current suppression in the right bias-triangles as compared to the left ones in Fig.~\ref{fig2}(b,vii). The $(1,4)$ state, a full-shell in the right dot and a single electron in the left dot, cannot accommodate the maximally spin-polarized $(2,3)$ state with $\sigma_z=\pm3/2$, e.g. $(\mathrm{T^-_sS_v},\mathrm{\downarrow K^-\uparrow K^+\downarrow K^+})$. The spin blockade lifts (marked by the triangle) upon accessing the spin-polarized four-particle excited state $\downarrow\downarrow/\uparrow\uparrow$ [gray in Fig.~\ref{fig2}(a,iv)] in the right dot.

The nature of Pauli blockade at various charge transitions is summarized in Fig.~\ref{fig1}(a) and compared with the conventional case where levels are filled with alternating spins. The richer structure in BLG stems from (1) the additional valley degrees of freedom, and as a result (2) the single-dot two-particle ground state being a spin-triplet~\cite{annikaexcitedstates,Samuel2particles} instead of a spin-singlet. 

From the above discussion, we see that the maximally spin- (valley-) polarized [i.e., with maximum $|\sigma_z|$ ($|\tau_z|$)] double dot ground state of a charge configuration is responsible for spin (valley) blockade. The maximum $|\sigma_z|$ ($|\tau_z|$) can be found by simply summing the total spin (valley) quantum number of the left and right single dot ground states $\sigma_\mathrm{NL}+\sigma_\mathrm{NR}$ ($\tau_\mathrm{NL}+\tau_\mathrm{NR}$). We can therefore summarize the above discussion into a simpler rule [illustrated in Fig.~\ref{fig2}(c)]: If the total spin (valley) number of the double dot ground state of the final charge configuration $\sigma_f$ ($\tau_f$) is lower than that of the initial charge configuration $\sigma_i$ ($\tau_i$), then there are (at least) initial states with $\sigma_z=\pm \sigma_i$ ($\tau_z=\pm \tau_i$) that do not exist in the final ground states, and this transition is spin- (valley-) blocked.

To substantiate the proposed three- and four-particle states spectra in Fig.~\ref{fig2}(a) that form the basis of the double-dot Pauli blockade analysis, we present experimental data investigating the spectra in magnetic field with the single-dot Kondo effect~\cite{goldhaber1998kondo}, similar to the analysis performed in Ref.~\cite{AnnikaKondo}.

With $V_\mathrm{L}$ as the plunger gate voltage, we tune the system to a single hole dot (the electron and hole spectra in BLG have so far been found identical) with strong dot-lead tunnel coupling, allowing for the observation of the Kondo effect~\cite{AnnikaKondo,goldhaber1998kondo}[sketched in Fig.~\ref{figkondo}(a), see Supplemental Materials~S1~\cite{supp} for details]. 

\begin{figure*}
	\includegraphics[width=17.8cm]{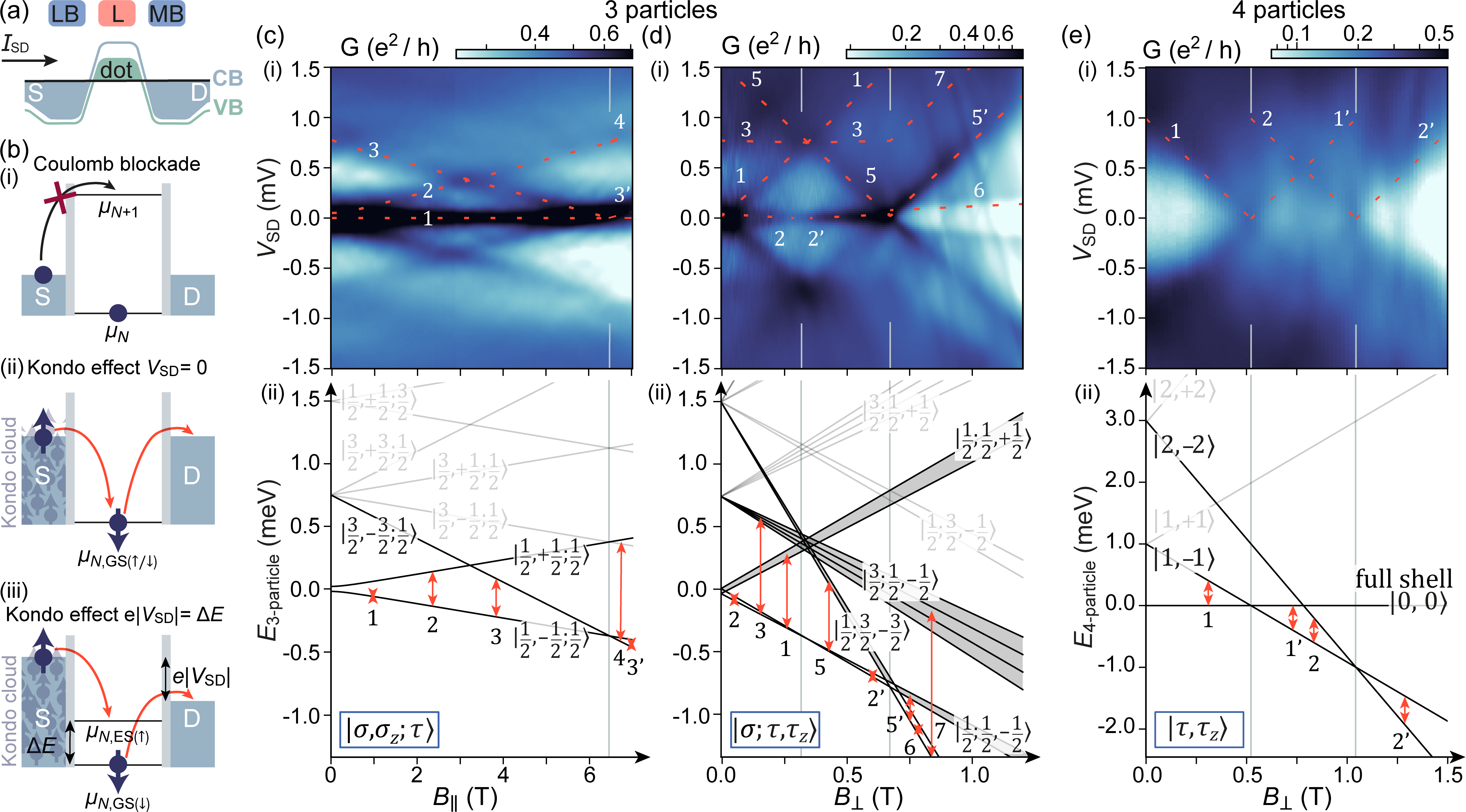}
	\caption{(a) Potential landscape along the channel for the single hole dot. (b) Schematics of (i) Coulomb blockade, (ii) 
    Kondo-assisted spin-flip co-tunnelings involving degenerate ground states with different quantum numbers at $V_\mathrm{SD}=0$, and (iii) involving the ground and an excited state at $V_\mathrm{SD}=\Delta E$, conserving energies. (c--e) (i) Evolution of conductance along finite bias cuts at fixed $V_\mathrm{L}$ for (c) three holes in in-plane field $B_\parallel$, (d) three holes in out-of-plane field $B_\perp$, and (e) four holes in out-of-plane field $B_\perp$, with (ii) respectively deduced three- and four-particle level spectra and relevant co-tunneling transitions
    . States are labeled with total spin (valley) numbers $\sigma$ ($\tau$), and projected spin (valley) numbers $\sigma_{z}$ ($\tau_z$).}
	\label{figkondo}
\end{figure*}

The Coulomb blockade in a QD [Fig.~\ref{figkondo}(b,i)] can be lifted by the Kondo effect [Fig.~\ref{figkondo}(b,ii)]~\cite{kondo1964resistance,wilson1975renormalization, goldhaber1998kondo, cronenwett1998tunable,van2000kondo, nygaard2000kondo}, allowing for finite conductance $G$: the unpaired spin (valley) state in the dot attracts a cloud of carriers with opposite spins (valleys) to the strongly coupled lead, assisting spin- (valley-) flip co-tunneling events~\cite{AnnikaKondo}. For energy conservation, the energy gained by the carrier escaping the dot equates to the energy lost by the carrier entering simultaneously from the lead. Therefore, a conductance resonance around zero $V_\mathrm{SD}$ indicates the existence of degenerate $N$-particle ground states with different spin (valley) quantum numbers---as observed in the Coulomb diamonds in Fig.~S1 for one, two, and three holes. The absence of this resonance for four holes confirms the formation of a singly degenerate full shell with paired spins and valleys. At finite $V_\mathrm{SD}$, co-tunnelings with excited states become possible [Fig.~\ref{figkondo}(b,iii)]. Conductance resonances occur when $e|V_\mathrm{SD}|=\mu_{N,ES}-\mu_{N,GS}:=\Delta E$, i.e. when the bias provides exactly the energy difference between the states. From the Coulomb diamonds in Fig.~S1 we obtain charging energies $\approx\SI{4}{meV}$, larger than the energy scales in the following discussions. 

Fixing $V_\mathrm{L}$ for three and four holes (See Supplemental Materials.~S1~\cite{supp} for the corresponding Coulomb diamonds and the chosen $V_\mathrm{L}$), we apply external in-plane $B_\parallel$ and out-of-plane $B_\perp$ field 
which shift the energies of spin and valley states.
Slopes of conductance resonances in $B$-fields offer information on the character of the states: a slope of $\pm g_\mathrm{s}$ ($\pm g_\mathrm{v}$), corresponds to a spin- (valley-) flip co-tunneling transition between the ground state and an excited state with different spin (valley) numbers $\Delta\sigma_z=\pm1$ ($\Delta\tau_z=\pm1$). For the discussion of the Kondo effect, we label the single QD states exclusively with spin and valley numbers for compactness.
We discuss for positive $V_\mathrm{SD}$ only, though the physics is equivalent for negative $V_\mathrm{SD}$.

With these rules in mind, we first look at conductance maps Fig.~\ref{figkondo}(c,i) for three particles in in-plane magnetic field, which couple only to spins but not to valleys. We thus see resonance 1 at zero-bias corresponding to valley-flip co-tunnelings, indicating the ground-state valley degeneracy $\tau_z=\pm1/2$. Resonance 2 splits off from 1 with slope $+g_\mathrm{s}\approx2$, indicating spin-flip co-tunnelings and thus the zero-field spin degeneracy of the ground state $\sigma_z=\pm1/2$. 
Starting from $\SI{0.7}{mV}$ at $B_\parallel=0$, resonance 3 shows a slope of $-g_\mathrm{s}\approx2$, indicating spin-flip co-tunnelings between the ground state with $\sigma_z=-1/2$ and an excited state with lower spin number. This maximally spin-polarized three-particle state with $\sigma_z=-3/2$ appears as an excited state at $\SI{0.7}{meV}$ as it requires contributions of an excited orbital level. 
This state is also responsible for lifting the $(1,2)\to(0,3)$ and $(2,2)\to(1,3)$ spin blockade [Fig.~\ref{fig2}(b,iii and vi), squares, also sketched in Fig.~\ref{fig2}(a,iii)]. This excited state becomes the ground state at $B_\parallel>\SI{7}{T}$. Co-tunneling events then arise from this state.

In out-of-plane field [Fig.~\ref{figkondo}(d)], as $g_\mathrm{v}\gg g_\mathrm{s}$ we resolve mainly valley splitting. The non-resolved spin-multiplet bundles are colored gray together. Being a zero-bias resonance in $B_\parallel$, resonance 1 splits off in $B_\perp$ with $g_\mathrm{v}\approx40$, corresponding to valley-flip co-tunnelings. We observe a `focusing' effect of resonance 2 becoming the narrowest at $\SI{0.35}{}\pm\SI{0.10}{T}$, arising from the competition between the zero-field splitting $\Delta_\mathrm{SO}$~\cite{AnnikaKondo,Luca1001} and the Zeeman splitting, changing the ground state from $\sigma_z=1/2;\tau_z=-1/2$ to $\sigma_z=-1/2;\tau_z=-1/2$. 
With $g_\mathrm{s}=2$ we estimate $\Delta_\mathrm{SO}\approx\SI{40}{}\pm\SI{10}{\mu eV}$, slightly lower than that reported for one-particle states~\cite{AnnikaKondo,Luca1001,Leakage}. 
Starting from $\SI{1.5}{mV}$ at $B_\perp=0$, resonance 5 with $g\sim40$ corresponds to valley-flip co-tunnelings between the ground state with $\tau_z=-1/2$ and an excited state with lower valley number $\tau_z=-3/2$. 
Since there exists no resonance with a slope of $g\sim-40$ starting from $\SI{0.7}{mV}$ at $B_\perp=0$, we conclude that the $\sigma_z=-3/2$ excited states contain no $\tau_z=-3/2$ states.
At $B_\perp>\SI{0.7}{T}$, the valley-polarized $\tau_z=-3/2$ state becomes the ground state. From there resonance 6 with $g_s=+2$ corresponds to spin-flip co-tunnelings between the $\tau_z=-3/2;\sigma_z\pm1/2$ states. Resonances 5', and 7 with slopes $g\sim+40$ correspond to valley-flip co-tunnelings of the $\tau_z=-3/2$ ground state
with the spin-doublet of $\sigma=1/2;\tau=1/2,\tau_z=-1/2$, and with the quadruplet bundle of $\sigma=3/2;\tau=1/2,\tau_z=-1/2$, respectively. 

It is worth noting that the three-particle $\sigma_z=3/2$ state is lower in energy ($\sim\SI{0.7}{meV}$) than the $\tau_z=3/2$ state ($\sim\SI{1.5}{meV}$), similar to the two-particle case~\cite{annikaexcitedstates,Samuel2particles} where, counter-intuitively, the spin-triplet ($\sigma=1$) is the ground state while the valley-triplet ($\tau=1$) is the excited state.

We now turn our attention to four-particle states in out-of-plane field. Intuitively, the four-hole ground state is a full shell with degeneracy of 1, and indeed no zero-bias peak occurs at zero field. Starting at $\SI{1}{mV}$, resonance 1 shows $g_\mathrm{v}\sim-40$, corresponding to valley-flip co-tunnelings between the full shell and an excited state with $\tau_z=-1$. This state becomes the ground state at $B_\perp>\SI{0.5}{T}$
. At $B_\perp>\SI{1.1}{T}$, a fully valley-polarized state with $\tau_z=-2$ becomes the ground state. Resonances 2 and 2' arise from co-tunnelings between the $\tau_z=-2$ and the $\tau_z=-1$ states. Tracing resonance 2 back to $B_\perp=0$, we obtain energy of the $\tau_z=-2$ excited state $\approx\SI{3}{meV}$.

The excited states are higher in energy due to involvement of higher orbital levels. Their energy deduced from the Kondo measurements agree well with that deduced from lifting of the double-dot blockade (Fig.~\ref{fig2}), which also agrees qualitatively with the calculated orbital levels~\cite{AngelikaQuartetStates,Knothe2022}.

As the ground states change in $B$-field, 
the double-dot Pauli blockade also changes in nature.
For charge degeneracies $(1,2)$--$(0,3)$, $(1,3)$--$(0,4)$, and $(2,2)$--$(1,3)$, we
study conductance along the detuning-axis of bias-triangles in $B_\perp$ (see Supplemental Materials~S2~\cite{supp}). The magneto-spectroscopy shows the evolution of Pauli blockade in $B_\perp$, and 
agrees with the suggested [Fig.~\ref{fig2}(a)] and demonstrated [Fig.~\ref{figkondo}(c--e)] $N$-particle state spectra.

To conclude, we established three- and four-particle state spectra in BLG QDs by examining their magnetic field dependence with the Kondo effect
. By discussing the relevant $N$-particle states and inspecting the double-dot bias-triangle measurements between a multitude of charge configurations, we reveal the Pauli blockade catalogue for BLG QDs as summarized in Fig.~\ref{fig1}(c,i), notably richer and intriguingly different compared to the canonical spin-blockade~\cite{JohnsonBlockade} [Fig.~\ref{fig1}(c,ii)]. The rich BLG state spectra and Pauli blockade catalogue offer an abundance of novel qubit operational positions, allowing for improved qubit manipulation and control exploiting unique spin and valley properties.


\section*{acknowledgments}

We are grateful for the technical support of Peter M\"arki and Thomas B\"ahler. We acknowledge financial support by the European Graphene Flagship Core3 Project, H2020 European Research Council (ERC) Synergy Grant under Grant Agreement 951541, the European Union’s Horizon 2020 research and innovation programme under grant agreement number 862660/QUANTUM E LEAPS, NCCR QSIT (Swiss National Science Foundation, grant number 51NF40-185902). R. G.  acknowledges funding from the European Union’s Horizon 2020 research and innovation programme under the Marie Skłodowska-Curie Grant Agreement No. 766025. K.W. and T.T. acknowledge support from the Elemental Strategy Initiative conducted by the MEXT, Japan, Grant Number JPMXP0112101001, JSPS KAKENHI Grant Number 19H05790 and JP20H00354.

\section*{Data availability}
The data supporting the findings of this study is made available via the ETH Research Collection.

\section*{Competing interests}
The authors declare no competing interests.

\nocite{Hiske2018electrostatically}\nocite{wang2013drytransferedge}

%

\clearpage
\newpage

\setcounter{section}{0} 
\renewcommand\thesection{S~\arabic{section}} 

\setcounter{figure}{0}
\renewcommand\thefigure{S\arabic{figure}}

\section{S1.~Methods}
\label{section:methods}
The device is fabricated as described in Ref.~\cite{Hiske2018electrostatically, MariusPRX, Blockade,Tunabledot,Leakage}. A false-colored AFM image of the sample is shown in Fig.~1(b,i). Stacked with the dry-transfer technique~\cite{wang2013drytransferedge}, the van der Waals heterostructure lies on a silicon chip with $\SI{285}{nm}$ surface SiO$_2$. The stack consists of a bottom graphite back gate, and on top of it a BLG flake encapsulated in $\SI{38}{nm}$ thick bottom and $\SI{20}{nm}$ thick top hBN flakes. Ohmic edge contacts with Cr and Au of $\SI{10}{}$ and $\SI{60}{nm}$ thickness, respectively, are evaporated after etching through the top hBN flake with reactive ion etching. A pair of $\SI{5}{nm}$ thick Cr, $\SI{20}{nm}$ thick Au split gates [gray in Fig.~1(b,i)] are deposited on top, defining a $\SI{1}{\micro m}$ long, $\SI{100}{nm}$ wide channel. Separated by a layer of $\SI{30}{nm}$ thick amorphous Al$_2$O$_3$ grown by atomic layer deposition, finger gates [red and blue in Fig.~1(b,i)] of $\SI{20}{nm}$ in width, and $\SI{5}{nm}$ Cr and $\SI{20}{nm}$ Au in thickness, lie across the channel. Neighboring finger gates are separated by $\SI{75}{nm}$ from center to center.

In BLG, a band-gap is formed near the $K^\pm$ valleys when applying a displacement field perpendicular to the sheet of BLG \cite{Ohta2006BLGband,mccann2006BLGband,Oostinga2008BLG}, where the size of the gap increases with the strength of the displacement field. With dual-gating, we have control over both the size of the gap, and the doping, in the gated region. 

Our double electron quantum dots studied in the first part of this work are defined electrostatically in the same BLG device with the method described in Refs.~\cite{Blockade, Leakage}. We apply a positive global graphite back-gate voltage $V_\mathrm{BG}=\SI{5}{V}$, tuning the whole sheet of BLG into an $n$-doped regime. With negative voltages $V_\mathrm{SG}=\SI{-3.645}{V}, \SI{-3.53}{V}$ applied to the split gates, we open up a band gap underneath the split gates, and simultaneously tune the Fermi energy $E_\mathrm{F}$ into the middle of the gap. Thus we confine the current flow $I_\mathrm{SD}$ to be only along this $n$-type 1D-channel. With another layer of gates deposited on top, we gain control locally over the potential landscape within this channel. As shown in Fig.~1(b,ii), we use the three barrier gates (blue) LB, MB, and RB to tune the region underneath them into the gap, forming our left, middle, and right barriers. The barrier gate voltages control the location of the Fermi energy $E_\mathrm{F}$ in the gap, and hence the strength of the barriers. In general, here the tunnel coupling decreases with more negative barrier gate voltages, until the voltages are negative enough for the formation of $p$-type dots underneath the barriers. The gates L and R serve as plunger gates for the electron dot L and R, respectively. With more negative plunger gate voltages $V_\mathrm{L}$ and $V_\mathrm{R}$, we can deplete the electron dots cleanly to the last carrier. 

Our single-hole quantum dot, strongly coupled to the leads studied in the second part of this work, is defined electrostatically in the same BLG device with the method described in Refs.~\cite{Tunabledot, AnnikaKondo}. We apply a positive global graphite back-gate voltage $V_\mathrm{BG}=\SI{3.7}{V}$, tuning the whole sheet of BLG into an $n$-doped regime. With negative voltages $V_\mathrm{SG}=\SI{-2.85}{V}$ applied to the split gates, we open up a bandgap underneath the split gates, and simultaneously tune the Fermi energy $E_\mathrm{F}$ into the middle of the gap. Thus we confine the current flow $I_\mathrm{SD}$ to be only along this $n$-type 1D-channel. With another layer of gates deposited on top, we gain control locally over the potential landscape within this channel. We use the plunger gate voltage $V_\mathrm{L}$ to tune the region beneath gate L into a $p$-type regime. The naturally forming $p$-$n$ junctions, therefore, become tunnel barriers of this single hole dot formed underneath gate L. With barrier gates LB and MB we can shift the bands underneath them in energy, and as a result, tune the width of the $p$-$n$ junction and hence the tunnel coupling~\cite{Tunabledot}. More positive barrier gate voltages give rise to narrower $p$-$n$ junctions, and hence a larger and more strongly coupled dot to the leads. For the measurements demonstrated in this work, the $V_\mathrm{LB}=V_\mathrm{MB}$ is set to $\SI{5.6655}{V}$ for Fig.~3(c), $\SI{5.6702}{V}$ for (d), and $\SI{4.8141}{V}$ for (e). 

\begin{figure}
	\includegraphics[width=8.5cm]{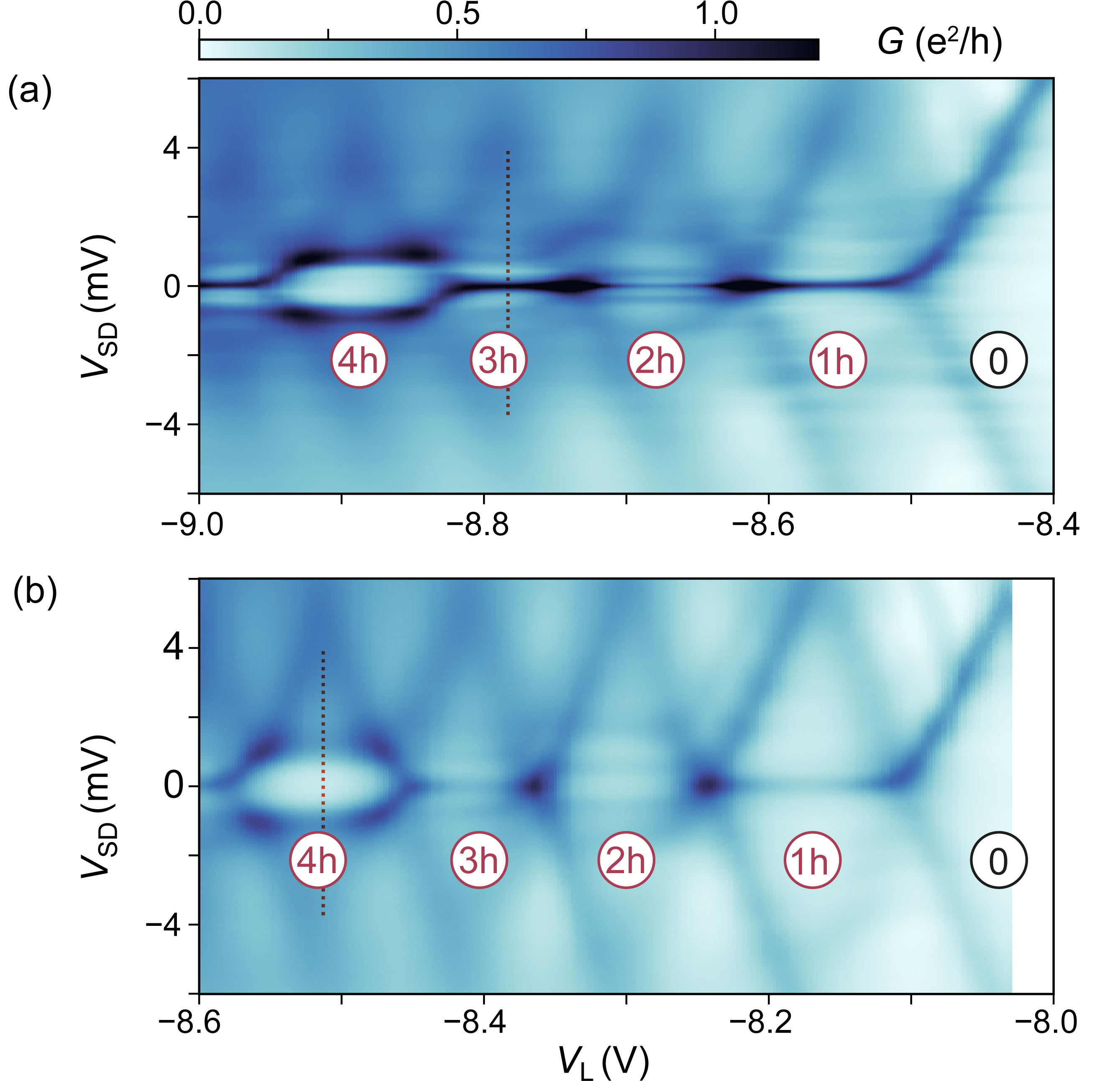}
	\caption{Coulomb diamonds near the regime of the Kondo effect measurements for (a) three-particle measurements shown in Fig.~3(c,d), and (b) four-particle measurements shown in Fig.~3(e). Stable hole occupancies are labelled in the diamonds. The red dotted lines mark the plunger gate voltages at which the three- and four-hole measurements shown in Fig.~3 in the main text are taken.} 
	\label{figkondodiamond}
\end{figure}

The plunger gate voltage $V_\mathrm{L}$ is set to be in the Coulomb blockaded region for three holes in (c) $V_\mathrm{L}=\SI{-8.77}{V}$, and in (d) $V_\mathrm{L}=\SI{-8.78}{V}$, and for four holes in (e) $V_\mathrm{L}=\SI{-8.515}{V}$. They are chosen to be in the middle of the Coulomb blockade diamond with the respective number of holes, as shown in Fig.~\ref{figkondodiamond} (red dotted lines).

The measurements are performed in a dilution refrigerator with base temperature of $\sim\SI{100}{\milli K}$. 

\section{S2.~Magnetospectroscopy along double-dot detuning-axis}

\label{section:det}
\begin{figure*}
	\includegraphics[width=17.8cm]{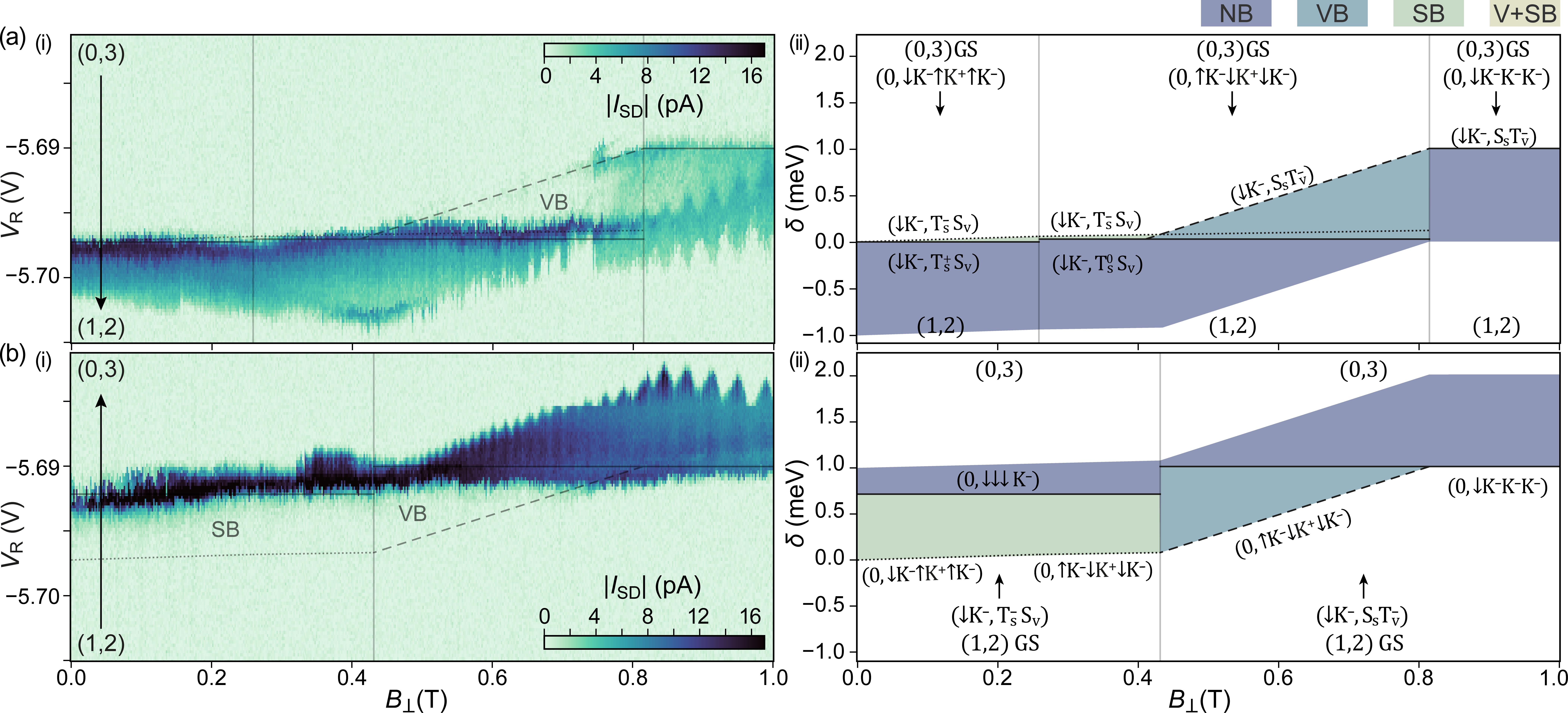}
	\caption{Evolution in the magnetic field of transitions for $V_\mathrm{SD}=$ (a) $\SI{-1}{mV}$ for transitions $(0,3)\to(1,2)$, and (b) $\SI{1}{mV}$ for transitions $(1,2)\to(0,3)$. (i) Line-cuts along the $\delta$-axis (plotted as $V_\mathrm{R}$). (ii) Calculated transitions with $\Delta_\mathrm{SO}=\SI{30}{\micro eV}$, excited state energy at zero-field measured from the zero-field ground state $\Delta E=\SI{0.7}{meV}$ for $\downarrow\downarrow\downarrow K^-_{N=3}$, $\Delta E=\SI{2.0}{meV}$ for $\downarrow K^-K^-K^-_{N=3}$, $\Delta E=\SI{1.0}{meV}$ for $\mathrm{S_sT^-_v}_{N=2}$, $g_\mathrm{v}=42$ and $g_\mathrm{s}=2$. Purple, blue, green, and yellow represent non-blocked (NB), valley-blocked (VB), spin-blocked (SB), and valley-and-spin-blocked (V+SB) regions, respectively, with corresponding transitions sketched in solid, dashed, dotted, and dash-dotted lines.}
	\label{fig1203}
\end{figure*}

\begin{figure*}
	\includegraphics[width=17.8cm]{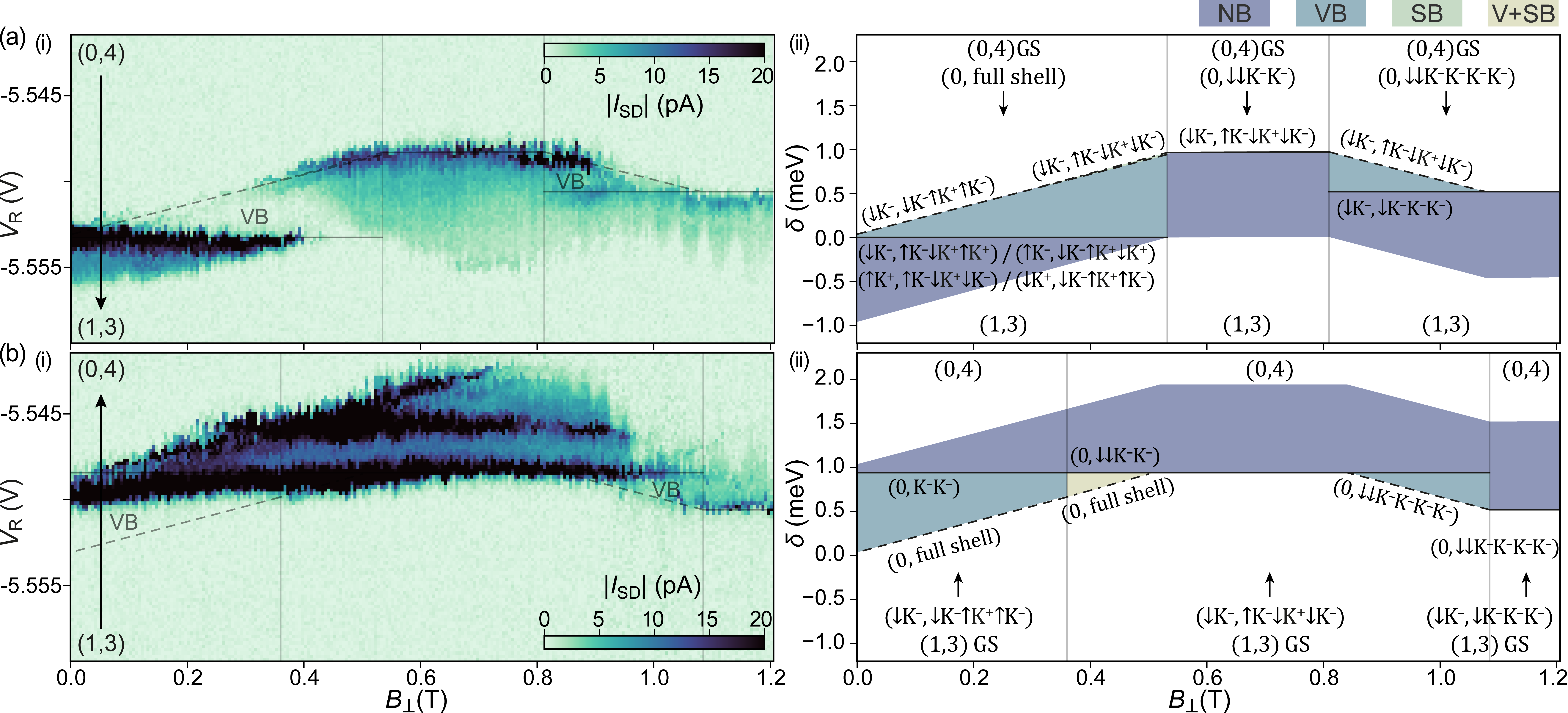}
	\caption{Evolution in the magnetic field of transitions for $V_\mathrm{SD}=$ (a) $\SI{-1}{mV}$ for transitions $(0,4)\to(1,3)$, and (b) $\SI{1}{mV}$ for transitions $(1,3)\to(0,4)$. (i) Line-cuts along the $\delta$-axis (plotted as $V_\mathrm{R}$). (ii) Calculated transitions with $\Delta_\mathrm{SO}=\SI{40}{\micro eV}$, excited state energy at zero-field measured from the zero-field ground state $\Delta E=\SI{0.75}{meV}$ for $\downarrow\downarrow\downarrow K^-_{N=3}$, $\Delta E=\SI{1.9}{meV}$ for $\downarrow K^-K^-K^-_{N=3}$, $\Delta E=\SI{0.9}{meV}$ for $\downarrow\downarrow K^-K^-_{N=4}$, $\Delta E=\SI{2.4}{meV}$ for $\downarrow\downarrow K^-K^-K^-K^-_{N=4}$, $g_\mathrm{v}=30$ and $g_\mathrm{s}=2$. Purple, blue, green, and yellow represent non-blocked (NB), valley-blocked (VB), spin-blocked (SB), and valley-and-spin-blocked (V+SB) regions, respectively, with corresponding transitions sketched in solid, dashed, dotted, and dash-dotted lines.}
	\label{fig1304}
\end{figure*}

\begin{figure*}
	\includegraphics[width=17.8cm]{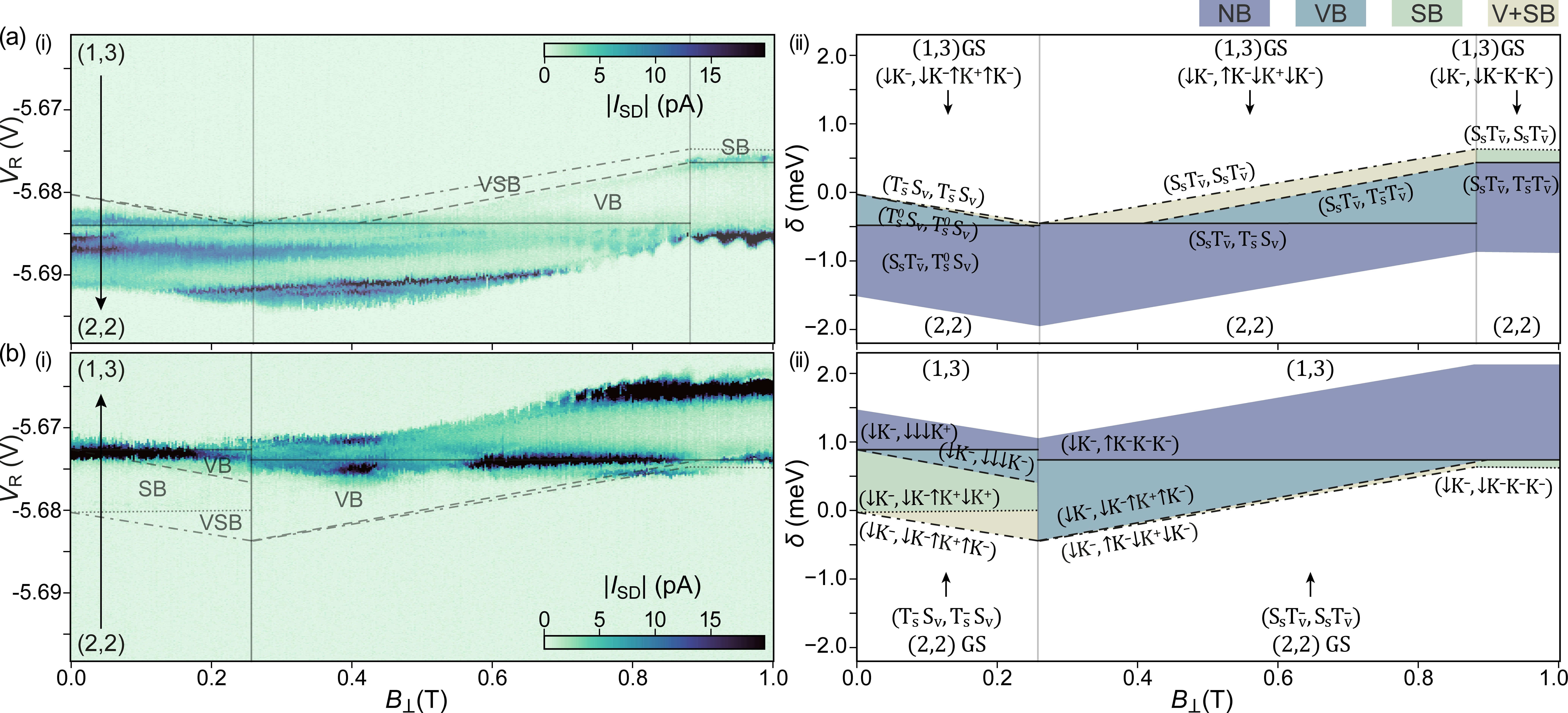}
	\caption{Evolution in the magnetic field of transitions for $V_\mathrm{SD}=$ (a) $\SI{-1.5}{mV}$ for transitions $(1,3)\to(2,2)$, and (b) $\SI{1.5}{mV}$ for transitions $(2,2)\to(1,3)$. (i) Line-cuts along the $\delta$-axis (plotted as $V_\mathrm{R}$). (ii) Calculated transitions with $\Delta_\mathrm{SO}=\SI{30}{\micro eV}$, excited state energy at zero-field measured from the zero-field ground state $\Delta E=\SI{0.9}{meV}$ for $\downarrow\downarrow\downarrow K^-_{N=3}$, $\Delta E=\SI{1.65}{meV}$ for $\downarrow K^-K^-K^-_{N=3}$, $\Delta E=\SI{0.45}{meV}$ for $\mathrm{S_sT^-_v}_{N=2}$, $\Delta E=\SI{0.3}{meV}$ for $\mathrm{T^-_sT^-_v}_{N=2}$, $g_\mathrm{v}=32$ and $g_\mathrm{s}=2$. Purple, blue, green, and yellow represent non-blocked (NB), valley-blocked (VB), spin-blocked (SB), and valley-and-spin-blocked (V+SB) regions, respectively, with corresponding transitions sketched in solid, dashed, dotted, and dash-dotted lines.}
	\label{fig2213}
\end{figure*}

In this section we show magnetospectroscopy measurements along the detuning-axis of the double dot bias triangles. The line cuts were always taken from the base to the tip of the triangle for the lower one of the pair of bias triangles (electron transport cycle). We apply an out-of-plane magnetic field $B_\perp$ and examine the evolution of the conductance along the $\delta$-axis. The resulting detuning (plotted as $V_\mathrm{R}$) v.s. $B_\perp$ maps are shown in Fig.~\ref{fig1203}(i), Fig.~\ref{fig1304}(i), and Fig.~\ref{fig2213}(i), for $(1,2)$--$(0,3)$, $(1,3)$--$(0,4)$, and $(2,2)$--$(1,3)$ charge degeneracies, respectively, where in (a) we apply negative bias $V_\mathrm{SD}$ such that the electrons travel from right to the left dot, and in (b) we apply positive bias $V_\mathrm{SD}$, and electrons travel from left to the right dot. The corresponding calculated transitions are labeled with initial and final states in (ii), with non-blocked (NB), valley-blocked (VB), spin-blocked (SB), and spin-and-valley-blocked (V+SB) colored in purple, blue, green, and yellow, respectively. The $\delta$ here is calculated from the plunger gate voltages $V_\mathrm{R}$ and $V_\mathrm{L}$. Relevant non-blocked, valley-blocked, spin-blocked, and spin-and-valley-blocked transitions, are sketched in straight, dashed, dotted, and dash-dotted lines, respectively. The parameters for the state energy calculations are listed in the respective figure captions. The analysis follows closely to what has been performed in Ref.~\cite{Blockade} around the $(1,1)$--$(0,2)$ charge degeneracy. 

When $B_\perp$ is applied, the states shift in energies due to spins and valleys coupling to $B_\perp$ with $g_\mathrm{s}$ and $g_\mathrm{v}$. Spin (valley) blocked transitions require spin (valley) flips, hence they show up in the $\delta$ v.s. $B_\perp$ measurements as lines with slopes of $g_\mathrm{s}$ or $g_\mathrm{v}$. Note, however, that we have for simplicity assumed that the left and the right dots are the same, i.e., the valley $g$-factor $g_\mathrm{v}$ is the same, and the same excited states occur at the same energies. This is not necessarily true as both of these quantities depend upon the exact dot geometry and potential landscape~\cite{Tunabledot,Samuel2particles}, though they are rather similar due to the similar environments. As a result of the differences, a non-blocked transition can also have a finite slope, and a spin-blocked transition can have a slope that is not exactly $g_\mathrm{s}$, arising from the difference of valley $g$-factor $\Delta g_\mathrm{v}$ in the two dots. 

The baseline of the bias triangles at zero $B_\perp$ is defined as zero-detuning $\delta=0$, when the electrochemical potential of the ground states of the initial and the final charge configuration aligns. If this ground-state to ground-state transition is prohibited by Pauli blockade, then current is suppressed from this transition onward, until meeting an excited state of the final charge configuration where the spin and valley quantum numbers match, lifting the blockade [see Fig.~2(c)]. A kink of the baseline (and hence a kink towards the edge of the bias window at large detuning $\delta$) therefore indicates a change in either of the ground states of the initial and final charge configuration --- the ground states change because a more spin or valley polarized state is lowered in energy by $B_\perp$ enough to overcome its excited state energy at zero-field $\Delta E$. Since the existence of Pauli blockade is always defined by the possible transitions initiated from the ground state of the initial charge configuration, we use vertical lines to separate regions with these different ground states. We characterize these transitions by the simple single-particle state spectrum picture discussed in Fig.~2(a), with experimentally demonstrated one-particle spectrum in Refs.~\cite{AnnikaKondo,Luca1001}, two-particle spectrum in Ref.~\cite{annikaexcitedstates,AnnikaKondo}, and with the three- and four-particle state spectra discussed in Fig.~3. We label the ground state to ground state transitions [the ones closest to the initial charge configurations, so the highest transitions for panels (a) with a negative bias, and the lowest transitions for panels (b) with positive bias], and if this transition is blocked, we label the next available transitions involving excited states in the final charge configuration that would lift this blockade. 

For three and four particle states involving excited orbital levels, for simplicity we label these states with only the unpaired spin and valley numbers, e.g., $\downarrow K^-\downarrow K^+\downarrow K^-_{N=3}$ becomes $\downarrow\downarrow\downarrow K^-_{N=3}$. To avoid confusion between three- and one-particle states and for consistency with the labeling in Fig.~2(a), we maintain the labeling for the three particle ground states: the first two particles have paired spin and valleys, and the last particle characterizes the unpaired spin and valley, e.g., $\downarrow K^-\uparrow K^+ \uparrow K^-_{N=3}$. Note also, that we transfer from labeling the three and four particle states with their quantum numbers in Fig.~3 to writing directly the spin and valley projections. For example, the state labeled $\downarrow\downarrow\downarrow K^-_{N=3}$ here, was labeled $\ket{\sigma_z=-3/2;\tau_z=-1/2}$ in Fig.~3.

\subsection{(1,2)--(0,3) charge degeneracy}

For $(0,3)\to(1,2)$, the low-magnetic field $(0,3)$ ground state is $(0,\downarrow K^-\uparrow K^+ \uparrow K^-)$, so the ground-state to ground-state transition to the $(1,2)$ ground state $(\downarrow K^-,\mathrm{T^-_sS_v})$ is spin-blocked. 
However, this blockade is quickly lifted in energy within a Zeeman splitting by accessing the $(1,2)$ excited state $(\downarrow K^-,\mathrm{T^+_sS_v})$. 
At $B_\perp>\SI{0.25}{T}$ the $(0,3)$ ground state changes to $(0,\uparrow K^- \downarrow K^+ \downarrow K^-)$, but transition to $(1,2)$ ground state remains spin-blocked, lifted by the Zeeman split $(1,2)$ excited state $(\downarrow K^-,T^0_sS_v)$. 
At large enough $B_\perp>\SI{0.45}{T}$ the $(1,2)$ ground state changes to $(\downarrow K^-, \mathrm{S_sT^-_v})$, so the transition to it is now valley-blocked, until the same $(1,2)$ excited state as before lifts this blockade. We thus see a region of suppressed conductance, labeled VB. 
At $B_\perp>\SI{0.85}{T}$ the $(0,3)$ ground state becomes the valley-polarized $(0,\downarrow K^-K^-K^-)$, so transition to the $(1,2)$ ground state $(\downarrow K^-, \mathrm{S_sT^-_v})$ is allowed, and hence the current recovers. 

For $(1,2)\to(0,3)$, the low-magnetic field $(1,2)$ ground state is $(\downarrow K^-,\mathrm{T^-_sS_v})$, so to the $(0,3)$ ground state $(0,\downarrow K^-\uparrow K^+ \uparrow K^-)$ (or $(0,\uparrow K^- \downarrow K^+ \downarrow K^-)$ for $B_\perp>\SI{0.25}{T}$) we first encounter a spin-blocked region, lifted by the $(0,3)$ excited state $(0,\downarrow\downarrow\downarrow K^-)$ occurring at $\Delta E=\SI{0.7}{meV}$, agreeing with the $\Delta E$ observed in the three-particle Kondo measurement in Fig.~3(c). At high enough $B_\perp>\SI{0.45}{T}$ the $(1,2)$ ground state changes into $(\downarrow K^-, \mathrm{S_sT^-_v})$, so transition to the $(0,3)$ ground state $(0,\uparrow K^- \downarrow K^+ \downarrow K^-)$ now becomes valley-blocked, until the $(0,3)$ excited state $(0, \downarrow K^-K^-K^-)$ at $\Delta E=\SI{2.0}{meV}$ lifting the valley blockade. This excited state energy is also similar to that observed in Kondo measurement, keeping the order that the spin-polarized state is lower in energy than the valley-polarized state. At larger $B_\perp>\SI{0.85}{T}$, this valley-polarized three-particle excited state is lowered enough in energy to become the new three-particle ground state, and therefore the $(1,2)\to(0,3)$ transitions are non-blocked thereon.  

\subsection{(1,3)--(0,4) charge degeneracy}

For $(0,4)\to(1,3)$, we first start from the $(0,4)$ ground state being the full shell, so its transition to the $(1,3)$ ground state $(\downarrow K^-,\downarrow K^-\uparrow K^+ \uparrow K^-)$ (or $(\downarrow K^-,\uparrow K^- \downarrow K^+ \downarrow K^-)$ at $B_\perp>\SI{0.25}{T}$) is valley-blocked. This blockade is lifted when the valley-split $(1,3)$ states become available, such that the $(0,4)$ full shell can be split into corresponding $(1,3)$ states. At large enough $B_\perp>\SI{0.55}{T}$ the half valley-polarized $(0,4)$ state $(0,\downarrow\downarrow K^-K^-)$ is lowered enough in energy and becomes the ground state, transition to the $(1,3)$ ground state is therefore not blocked and current recovers over the entire $\SI{1}{meV}$ bias window. At even larger $B_\perp>\SI{0.8}{T}$ the fully valley-polarized $(0,4)$ state $(0, \downarrow\downarrow K^-K^-K^-K^-)$ becomes the ground state, hence transition to the $(1,3)$ ground state is valley blocked. Eventually the $(1,3)$ ground state also changes to the valley-polarized $(\downarrow K^-, \downarrow K^-K^-K^-)$, fully lifting the blockade and recovering the current. The locations of the kinks in $B_\perp$ marking the changes of four-particle ground states, agree well with that shown by the four-particle Kondo effect in Fig.~3(d). 

For $(1,3)\to(0,4)$, we begin first with the $(1,3)$ ground state $(\downarrow K^-, \downarrow K^-\uparrow K^+ \uparrow K^-)$ such that transition to the $(0,4)$ ground state of a full shell in the right dot is valley-blocked, until meeting the $(0,4)$ half valley-polarized excited state $(0,K^-K^-)$. When the $(1,3)$ ground state changes to $(\downarrow K^-, \uparrow K^-\downarrow K^+\downarrow K^-)$ at $B_\perp>\SI{0.35}{T}$, the ground-state to ground-state transition is spin and valley blocked, lifted by a similar $(0,4)$ excited state $(0,\downarrow\downarrow K^-K^-)$. This half valley-polarized excited state exists at an energy of $\Delta E\approx\SI{1}{meV}$, agreeing with that observed in the four-particle Kondo effect in Fig.~3(d). Another horizontal resonance appears in the non-blocked region at even higher energies, corresponding to another $(0,\downarrow\downarrow K^-K^-)$ state with an excited orbital state. The excited orbital state is $\approx\SI{0.5}{meV}$ higher in energy than the lower orbital state with the same spin and valley configuration. Eventually, this half valley-polarized state $(0,\downarrow\downarrow K^-K^-)$ becomes the $(0,4)$ ground state and current recovers in the entire bias window. At higher $B_\perp>\SI{0.8}{T}$ the fully valley-polarized $(0,4)$ state becomes the new ground state, and transitions from the $(1,3)$ $(\downarrow K^-, \uparrow K^-\downarrow K^+\downarrow K^-)$ is again valley blocked, until at even higher $B_\perp>\SI{1.1}{T}$, where the $(1,3)$ ground state also changes to the valley polarized $(\downarrow K^-, \downarrow K^-K^-K^-)$, lifting the blockade and recovering the current. 

\subsection{(2,2)--(1,3) charge degeneracy}

For $(1,3)\to(2,2)$, we start with the $(1,3)$ ground state $(\downarrow K^-, \downarrow K^-\uparrow K^+\uparrow K^-)$, and its transition to the $(2,2)$ ground state $(\mathrm{T^-_sS_v}, \mathrm{T^-_sS_v})$ is spin and valley blocked. Only within a Zeeman splitting energy though, the $(2,2)$ excited state $(\mathrm{T^0_sS_v}, \mathrm{T^0_sS_v})$ will become accessible, such that we have only valley blockade. The blockade is completely lifted when the two-particle excited state in one of the two dots becomes available such that we have access to $(\mathrm{S_sT^-_v}, \mathrm{T^0_sS_v})$. Because the spin singlet-triplet energy splitting $E_\mathrm{ST}$ in the two dots can be slightly different (it depends on the dot geometry and band gap etc.~\cite{AngelikaQuartetStates,Knothe2022,Samuel2particles}, it can result in two roughly horizontal resonances separated in energy by $|E_\mathrm{ST,L}-E_\mathrm{ST,R}|$. At higher energy, we have access to the two-particle orbital anti-symmetric excited states: spin-triplets valley-triplets. These states are higher in energy by $\Delta_\mathrm{AS}$ from the spin-singlet valley-triplets. At detuning energies $E_\mathrm{ST,L/R}+\Delta_\mathrm{AS,L/R}$, we have four additional resonances corresponding to (1,3) ground state aligning with (2,2) states: ($\mathrm{T^0_sT^-_v, T_sS_v}$), ($\mathrm{S_sT^-_v, T_sT^0_v}$), ($\mathrm{T_sT^0_v, S_sT^-_v}$) and ($\mathrm{T_sS_v, T^0_sT^-_v}$). At even higher energies $E_\mathrm{ST,L/R}+\Delta_\mathrm{AS,L}+\Delta_\mathrm{AS,R}$, we have two additional horizontal resonances corresponding to alignment with (2,2) states with orbitally anti-symmetric states in both dots: $(\mathrm{T^0_sT^-_v, T_sT^0_v})$, $(\mathrm{T_sT^0_v, T^0_sT^-_v})$. From the measurement, we can estimate $E_\mathrm{ST}\approx\SI{0.5}{meV}$, $\Delta_\mathrm{AS}\approx\SI{0.5}{meV}$.

At large enough $B_\perp>\SI{0.25}{T}$, the $(1,3)$ ground state changes to $(\downarrow K^-,\uparrow K^-\downarrow K^+\downarrow K^-)$ in $B_\perp$, such that a different two-particle spin-triplet, namely $(\mathrm{S_sT^-_v}, \mathrm{T^-_sS_v})$, is required to lift the blockade. Meanwhile, the $(2,2)$ ground state also becomes $(\mathrm{S_sT^-_v}, \mathrm{S_sT^-_v})$, and, transition from the $(1,3)$ ground state $(\downarrow K^-, \uparrow K^-\downarrow K^+\downarrow K^-)$ is spin and valley blocked. This is partly lifted upon hitting a $(2,2)$ fully polarized excited state $\SI{0.3}{meV}$ above the spin-singlets, a reasonable size as compared to that discussed in Ref.~\cite{Samuel2particles}, such that we have only valley blockade and current is suppressed but finite. The blockade is fully lifted upon the $(2,2)$ excited state $(\mathrm{S_sT^-_v}, \mathrm{T^-_sS_v})$. At large $B_\perp>\SI{0.9}{T}$ the valley-polarized $(\downarrow K^-, \downarrow K^-K^-K^-)$ becomes the $(1,3)$ ground state, and transition from it to the $(2,2)$ ground state becomes valley-blocked, which is lifted by the spin-polarized valley-polarized two-particle state. 

For $(2,2)\to(1,3)$, we start with the $(2,2)$ ground state $(\mathrm{T^-_sS_v}, \mathrm{T^-_sS_v})$. The transition to the $(1,3)$ ground state $(\downarrow K^-,\downarrow K^-\uparrow K^+\uparrow K^-)$ is spin and valley blocked, which is partially lifted when hitting the $(1,3)$ excited state $(\downarrow K^-,\downarrow K^-\uparrow K^+\downarrow K+)$ where it is then only spin blocked. When hitting the spin-polarized $(1,3)$ excited state $(\downarrow K^-, \downarrow\downarrow\downarrow K^-)$ the transitions are then valley-blocked. Only when hitting the other valley branch of this spin-polarized $(1,3)$ state $(\downarrow K^-, \downarrow\downarrow\downarrow K^+)$ the transition is fully lifted, recovering the current. At $B_\perp>\SI{0.25}{T}$ the $(2,2)$ ground state changes into $(\mathrm{S_sT^-_v}, \mathrm{S_sT^-_v})$, hence transition to the $(1,3)$ ground state $(\downarrow K^-, \uparrow K^-\downarrow K^+\downarrow K^-)$ is spin and valley blocked, until hitting the $(1,3)$ $(\downarrow K^-,\downarrow K^-\uparrow K^+\uparrow K^-)$ state, thereon it is valley blocked only. The blockade is fully lifted eventually by the valley-polarized $(1,3)$ excited state $(\downarrow K^-, \uparrow K^-K^-K^-)$. Its spin-down branch $(\downarrow K^-, \downarrow K^-K^-K^-)$ eventually becomes the $(1,3)$ ground state at $B_\perp>\SI{0.9}{meV}$, and transition from the $(2,2)$ ground state to it is first spin-blocked, but then completely lifted by the spin up branch.

\end{document}